\pdfoutput=1
\documentclass[journal=jacsat,manuscript=article]{achemso}
\setkeys{acs}{articletitle = true}
\usepackage[version=3]{mhchem} 
\usepackage[usenames,dvipsnames]{color}

\author{Caio P. de Castro}
\affiliation{Instituto de F\'{\i}sica, Universidade Federal da Bahia,
   Campus Universit\'{a}rio da Federa\c c\~ao,
   Rua Bar\~{a}o de Jeremoabo s/n,
40170-115, Salvador, BA, Brazil}
\email{caioporto@ufba.br}
\author{Thiago A. de Assis}
\affiliation{Instituto de F\'{\i}sica, Universidade Federal da Bahia,
   Campus Universit\'{a}rio da Federa\c c\~ao,
   Rua Bar\~{a}o de Jeremoabo s/n,
40170-115, Salvador, BA, Brazil}
\email{thiagoaa@ufba.br}
\author{Roberto Rivelino}
\affiliation{Instituto de F\'{\i}sica, Universidade Federal da Bahia,
   Campus Universit\'{a}rio da Federa\c c\~ao,
   Rua Bar\~{a}o de Jeremoabo s/n,
40170-115, Salvador, BA, Brazil}
\email{rivelino@ufba.br}
\author{Fernando de B. Mota}
\affiliation{Instituto de F\'{\i}sica, Universidade Federal da Bahia,
   Campus Universit\'{a}rio da Federa\c c\~ao,
   Rua Bar\~{a}o de Jeremoabo s/n,
40170-115, Salvador, BA, Brazil}
\email{fbmota@ufba.br}
\author{Caio M. C. de Castilho}
\affiliation{Instituto de F\'{\i}sica, Universidade Federal da Bahia,
   Campus Universit\'{a}rio da Federa\c c\~ao,
   Rua Bar\~{a}o de Jeremoabo s/n,
40170-115, Salvador, BA, Brazil}
\alsoaffiliation{Centro Interdisciplinar em Energia e Ambiente, Universidade Federal da Bahia, Campus Universit\'{a}rio da Federa\c{c}\~{a}o, 40170-115 Salvador, BA, Brazil}
\alsoaffiliation{Instituto Nacional de Ci\^{e}ncia e Tecnologia em Energia e Ambiente - INCTE\&A, Universidade Federal da Bahia,
   Campus Universit\'{a}rio da Federa\c c\~ao,
   Rua Bar\~{a}o de Jeremoabo s/n,
40170-280, Salvador, BA, Brazil}
\email{caio@ufba.br}
\author{Richard G. Forbes}
\affiliation{Advanced Technology Institute \& Department of Electrical and Electronic Engineering, University of Surrey, Guildford, Surrey GU2 7XH, UK}
\email{r.forbes@trinity.cantab.net}
\title[An \textsf{achemso} demo]
  {Modeling the Field Emission Enhancement Factor for Capped Carbon Nanotubes using the Induced Electron Density}

\abbreviations{IR,NMR,UV}

\begin{document}

 \newpage
\begin{abstract}
In many field electron emission experiments on single-walled carbon nanotubes (SWCNTs), the SWCNT stands on one of two well-separated parallel plane plates, with a macroscopic field FM applied between them. For any given location "L" on the SWCNT surface, a field enhancement factor (FEF) is defined as $F_{\rm{L}}$/$F_{\rm{M}}$, where $F_{\rm{L}}$ is a local field defined at ``L". The best emission measurements from small-radii capped SWCNTs exhibit characteristic FEFs that are constant (i.e., independent of $F_{\rm{M}}$). This paper discusses how to retrieve this result in quantum-mechanical (as opposed to classical electrostatic) calculations. Density functional theory (DFT) is used to analyze the properties of two short, floating SWCNTS, capped at both ends, namely a (6,6) and a (10,0) structure. Both have effectively the same height ($\sim 5.46$ nm) and radius ($\sim 0.42$ nm). It is found that apex values of local induced FEF are similar for the two SWCNTs, are independent of $F_{\rm{M}}$, and are similar to FEF-values found from classical conductor models. It is suggested that these induced-FEF values relate to the SWCNT longitudinal system polarizabilities, which are presumed similar. The DFT calculations also generate ``real", as opposed to ``induced", potential-energy (PE) barriers for the two SWCNTs, for FM-values from 3 V/$\mu$m to 2 V/nm. PE profiles along the SWCNT axis and along a parallel ``observation line" through one of the topmost atoms are similar. At low macroscopic fields the details of barrier shape differ for the two SWCNT types. Even for $F_{\rm{M}}=0$, there are distinct PE structures present at the emitter apex (different for the two SWCNTs); this suggests the presence of structure-specific chemically induced charge transfers and related patch-field distributions.

\end{abstract}
 \vspace{0.5cm}
 \newpage
\section{Introduction}

Significant progress has been made in understanding how small nanostructures can generate field electron emission (FE) \cite{Cole2015chapter,nature2018}. Carbon nanotube (CNT)-based field emitters \cite{Buldum,Vicent,Pascale,Nanotech2016} are particularly effective because an electrically conducting pointed nanostructure enhances an externally applied macroscopic field, $F_{\rm{M}}$, thereby creating high local electrostatic fields in the vicinity of the nanostructure apex.  Thus, a local field enhancement factor (FEF), $\gamma(\mathbf{r})$, can be defined as $\gamma(\mathbf{r}) = F(\mathbf{r})/F_{\rm{M}}$, where $F(\mathbf{r})$ is a particular local field at a point in three-dimensional space. Because of the variable aspect ratio (ratio of longitudinal to lateral dimensions) of CNTs and, consequently, the tunability of their FEFs, they find applications in many areas ranging from  nanoelectronics \cite{Hyegi, Nguyen} to biological sensors \cite{Tans,Wong}.

Some CNT field electron emitters (see e.g., Refs. \cite{BonardPRL2002,Dean2000}) exhibit linear Fowler-Nordheim (FN) plots satisfying the Forbes orthodoxy test \cite{Forbes2013}. This result demonstrates that, for such emitters, there exists a \textit{characteristic} FEF, $\gamma_{\rm{C}}$, that is constant for a wide range of applied voltages and macroscopic electric fields $F_{\rm{M}}$. This further implies that (for any given value of $F_{\rm{M}}$) there exists an electron tunneling barrier that can be considered  \textit{characteristic} for the field emitting CNT.

The physical effect of applying the field $F_{\rm{M}}$ is to polarize the electron density of the CNT emitter, leading to local changes in electron density. In the context of electronic structure theory, the \textit{change} in electron density at any point is called the ``induced electron density" and is proportional to a quantity called the ``density response function" \cite{parr1994density}. This response function will be different for each system (i.e., each nanostructure) and is a functional of the ground-state electron-density distribution of the whole emitter, as  this distribution exists in the absence of the applied macroscopic field.

One way of representing the field enhancing properties of a nanostructure is to calculate the so-called ``local induced FEF" by considering the emitter's induced-charge distribution \cite{ForbesES,JPCC2019,deCastro2019jap}. Provided that it is reasonable to make the assumption that there are no significant changes in the positions of the atomic nuclei, this induced-charge distribution is given by the induced electron density. It follows that one way of calculating a characteristic FEF is to use the induced-charge distribution to calculate related induced-field and induced-FEF distributions. There remains an issue of deciding at what point in space the characteristic FEF should be defined: this is discussed below. As just indicated, one theoretical way of calculating the induced-charge distribution is to calculate induced electron densities. In the context of electronic structure theory, this can be done by using linear response theory to calculate the density response function (see chapter 9 in Ref.  \cite{parr1994density}).

However, some earlier theoretical treatments of CNT behavior, for example,  Ref. \cite{PRBLi2005}, have used a different definition of characteristic FEF. These treatments use electronic structure theory to calculate the \textit{total} polarized charge distribution, and then use this to define (what we have called \cite{JPCC2019}) \textit{real} local-field and local-FEF distributions. With this approach, their characteristic FEF is a value of real-FEF taken at some chosen location in space above the CNT apex.

For example, in Ref. \cite{PRBLi2005}, the authors calculated the characteristic FEF by using the total polarized charge distribution of a long capped single-walled CNT (SWCNT). One of their conclusions was that the value of their characteristic FEF would depend on $F_{\rm{M}}$ (see their Fig. 8(a)). However, a dependence of characteristic FEF on $F_{\rm{M}}$ is in conflict with some of the best experimental work on CNTs, for example that of Bonard \textit{et al.} \cite{BonardPRL2002}, which generated \textit{linear} Fowler-Nordheim (FN) plots. A linear FN plot implies the existence of a \textit{constant} FEF. The FEF values extracted from the experiments were reported to be consistent with values calculated by modeling techniques that consider the CNT as a classical conductor with the shape of a hemisphere on a cylindrical post (the so-called ``HCP model"), and then apply finite-element methods based on Laplace's equation \cite{Edgcombe2001,Edgcombe}. 

At the nanoscale, an interesting issue is to understand how the CNT chirality impacts its field emission properties. Chirality effects on FE current density from opened-end SWCNTs have been reported when using a tight-binding approximation \cite{APL2000china, APL2004china}, but possible effects on characteristic FEFs have not been investigated. We have investigated this issue, by using density functional theory (DFT) \cite{Jones2018} to calculate FEF values for two limiting classes of capped SWCNTs, with distinct chiral indexes, i.e., (6,6) and (10,0), but with the same total lengths and radii.

A \textit{linear polarization regime} is defined as a range of macroscopic field ($F_{\rm{M}}$) values in which the total dipole moment of the floating CNT (as resolved parallel to the CNT axis) is proportional to $F_{\rm{M}}$. The coefficient of proportionality is called here the longitudinal system polarizability \cite{JPCC2005}. For a very broad range of $F_{\rm{M}}$-values, from 3-10 V/${\mu}$m to 1-2 V/nm, as applied to these SWCNTs, we have established that these emitters exhibit a linear polarization regime, with similar (real-charge) longitudinal polarizabilities (despite the difference in chiral indexes).

In principle, characteristic-FEF values should have a link with longitudinal system polarizabilities. So we might expect that characteristic values of related ``real-FEFs" would be similar, independent of chirality. At present, problems in establishing a reliable method of defining real-FEFs make it difficult to confirm this. But we do find that ``induced FEFs" derived from the induced-charge and induced-field distributions are similar, despite differences in the considered chiral indexes. Notwithstanding this, differences have been observed in apparently equivalent ``real" potential-energy barriers for the two types of SWCNTs, particularly with regard to barrier height with respect to the Fermi level.

This work is organized as follows. We present our models for the field emitting SWCNTs and related computational details. We then discuss the results for the electronic structure of the different floating SWCNTs studied, and derived emitter properties, including characteristic values of local induced-FEF, found from the induced-charge distribution. We also discuss results found for the shapes of potential-energy barriers. Finally, we summarize our main conclusions.  

\section{Computational methods and systems}
\label{Method}

\begin{figure}[h!b]
\centering
\includegraphics [scale=0.3] {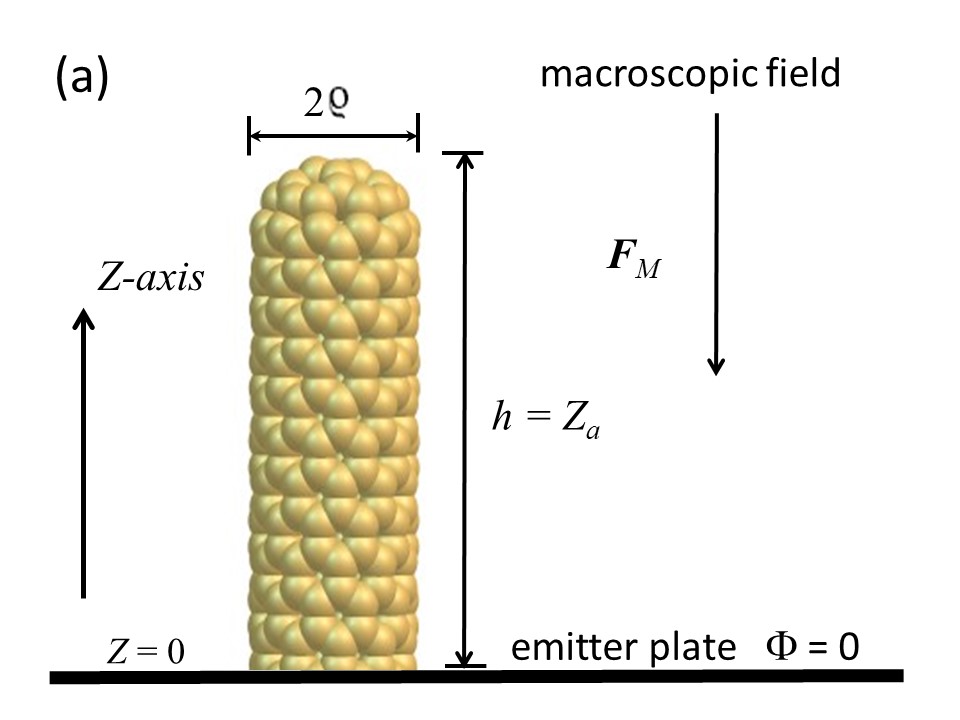}
\includegraphics [scale=0.3] {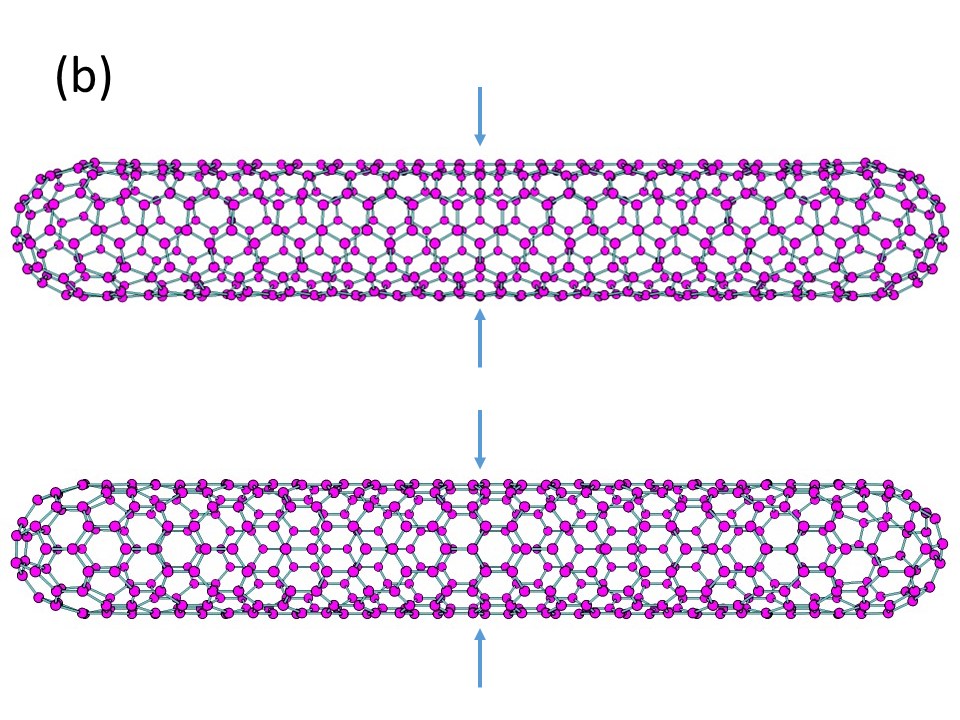}
\caption{(a) Representation of a field electron emission system equivalent to the top half of a floating SWCNT. The classical-conductor equivalent is the hemisphere-on-cylindrical-post (HCP) model. (b) Floating [top] (6,6) and [bottom] (10,0) SWCNTs used in this work. The arrows help ones to observe arm-chair and zig-zag structures in (6,6) and (10,0) SWCNTs, respectively.} 
\label{rep}
\end{figure}

The use of individual CNTs as functional electronic devices is of interest because they exhibit high sensitivity and fast response to electric fields \cite{Dai2002}. Theoretically, a floating capped SWCNT can be modelled quantum mechanically, and relevant results can be compared with the equivalent results obtained from the classical ``hemisphere-on-cylindrical-post (HCP)" model, where the post stands on a conducting emitter plate of large extent. In our case: (a) Laplace-type analysis, using finite element methods, is used to analyze the HCP model (e.g., Ref. \cite{JVSTB2019});and (b) density functional theory (DFT) \cite{Jones2018} is used for the quantum-mechanical analysis. The ``top half" of a simple model is represented schematically in Figure \ref{rep} (a). For the DFT calculations we selected (6,6) and (10,0) floating SWCNTs containing $540$ and $500$ carbon atoms, respectively [see Fig. \ref{rep} (b)]. These CNTs are both capped at both ends, and have approximately the same radius ($\varrho\approx 0.42$ nm) and total length ($2h \approx 5.46$ nm).

For these capped SWCNTs, first-principles calculations were carried out using DFT techniques as implemented in the SIESTA code \cite{Soler2002}. The structures were built in rectangular boxes with sizes 4 nm $\times$ 4 nm $\times$ 10.8 nm, and all carbon atoms were allowed to relax until atomic forces decrease below 0.05 eV/nm.  The lateral images in the box side-faces mean that the SWCNTs behave as if they were components in a regular square array of such emitters. This means that electrostatic-depolarization effects (sometimes called ``shielding" or ``screening") will be present, and the calculated apex-FEF values will be slightly smaller than those applicable to a totally isolated SWCNT. However, because the box lateral side-length is approximately  5 times the CNT radius, this depolarization effect is not expected to affect either the qualitative behavior or relative comparisons.

The Kohn-Sham equations were solved using the PBE exchange-correlation potential scheme \cite{PBE}, which has been demonstrated to work well for these kinds of systems \cite{RB,JPCC2019,deCastro2019jap}. Core electrons were described in terms of the Troullier-Martins norm-conserving pseudopotentials \cite{TM1991} and valence electrons with double-$\zeta$ basis sets, including polarization functions, using an energy cutoff of $300$ Ry and sampling in the $\Gamma$ point of the Brillouin-zone.

Using this approach, we performed DFT calculations in the presence of an applied external macroscopic field, $F_{\rm{M}}$, to generate the ``real" charge distributions of the SWCNTs, denoted here by $\rho_{\rm{r}}({\bf{r}},F_{\rm{M}})$. From these we generated the spatial distributions of the local ``real" electric field values $F_{\rm{r}}({\bf{r}},F_{\rm{M}})$, and determined the local ``real" FEF-values (LRFEFs), $\gamma_{\rm{r}}({\bf{r}},F_{\rm{M}})$, as defined by:

\begin{equation}
\gamma_r(\mathbf{r},F_{\rm{M}}) \equiv \frac{F_r(\mathbf{r},F_{\rm{M}})}{F_{\rm{M}}}.   
\label{Eqrealfef}
\end{equation}

In Eq.(\ref{Eqrealfef}), $\mathbf{r}$ is a position vector where a local FEF can be defined. The related ``induced-charge" distribution, $\rho_{\rm{i}}(\mathbf{r},F_{\rm{M}})$, is

\begin{equation}
\rho_{\rm{i}}(\mathbf{r},F_{\rm{M}}) \equiv \rho_{\rm{r}}(\mathbf{r},F_{\rm{M}}) - \rho_{\rm{r}}(\mathbf{r},0),
\label{indcharge}
\end{equation}
and can be used to define distributions in space of local induced-field values $F_{\rm{i}}(\mathbf{r},F_{\rm{M}})$ and related local induced FEF-values (LIFEFs), defined analogously to Eq. (\ref{Eqrealfef}) by

\begin{equation}
\gamma_{\rm{i}}(\mathbf{r},F_{\rm{M}})\equiv \frac{F_{\rm{i}}(\mathbf{r}, F_{\rm{M}})}{F_{\rm{M}}}.
\label{LIFEF}
\end{equation}
As we discuss below, this is an appropriate approach to determine FEF-values for nanosystems, taking into account the electronic structure of the material.

\section{Results and discussion}
\label{Results}	

\subsection{Density of states and electric response}
\begin{figure}[h!b]
\includegraphics [scale=0.5] {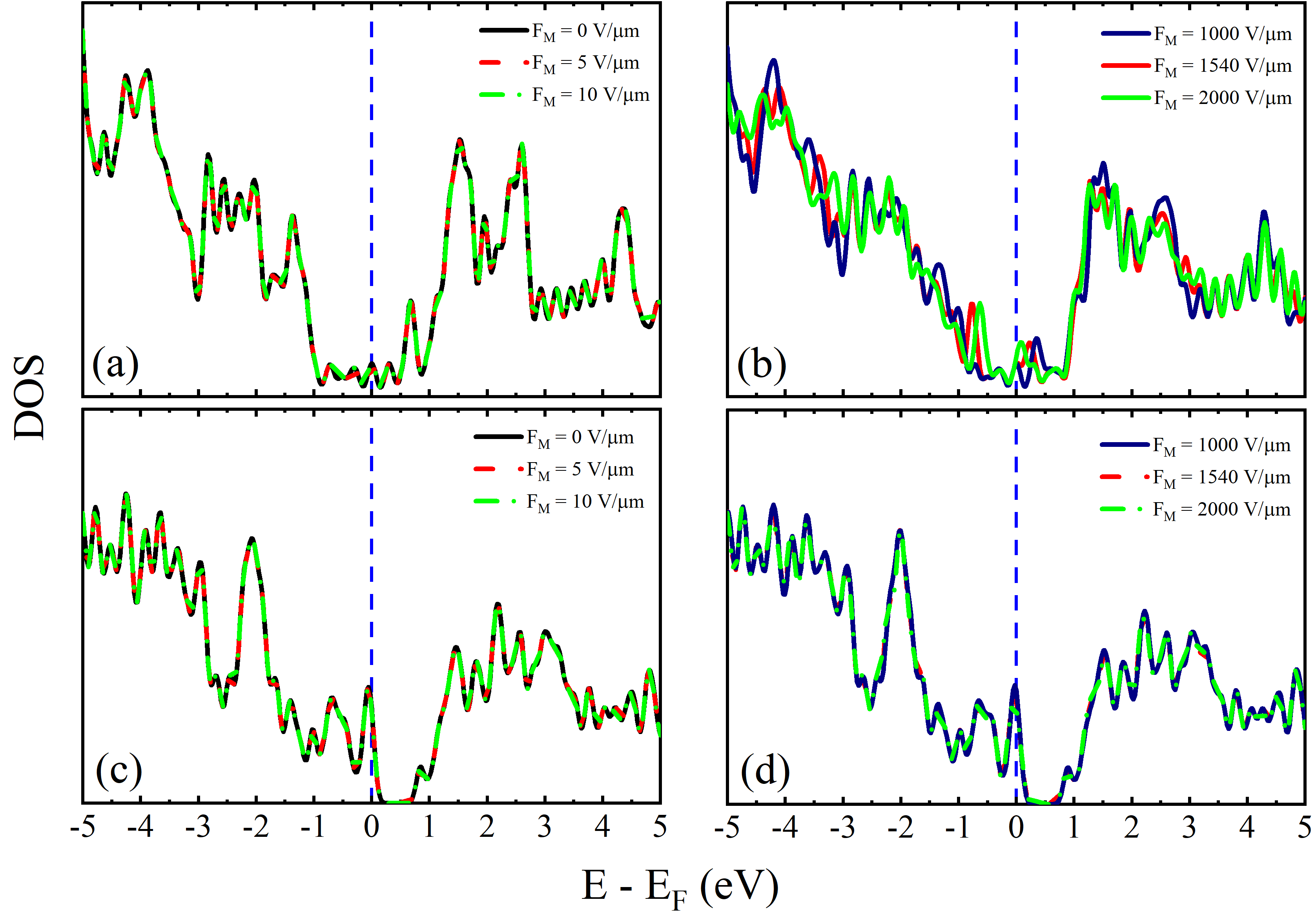}
\caption{Calculated density of states (DOS) for the two floating SWCNTs: (a) and (b) for (6,6), and (c) and (d) for (10,0), with macroscopic fields $F_{\rm{M}}$ = 0, 5, 10, 1000, 1540, and 2000 V/$\mu$m. The results for $F_{\rm{M}}$ = 0, 5, and 10 V/$\mu$m are presented in panels (a) and (c). The results for $F_{\rm{M}}$ = 1000, 1540, and 2000 V/$\mu$m are separately presented in diagrams (b) and (d).} \label{DOSCNT}
\end{figure}

Within the zone folding approximation \cite{Dresselhaus}, about 1/3 of possible SWCNTs exhibit a metallic or a quasi-metallic character, depending on the geometric structure, while all other 2/3 possible SWCNTs effectively have a gap in their densities of states (DOS) and are expected to behave much like semiconductors. Here, we consider two representative structures of these alternative cases, i.e., a (6,6) and a (10,0) SWCNT, but with both capped at both ends, in order to mimic the classical HCP model discussed above. This HCP model is often used to model/interpret experimental FE situations \cite{Edgcombe2001,Edgcombe,Xanthakis,Edgcombe,RBowring,ZENG2009,Unicamp2016,JVSTB2019}. We note that, although the ``infinite" (6,6) SWCNT is in principle gapless and the ``infinite" (10,0) SWCNT is in principle a semiconductor, the capping of the tubular structures results in a metallic character for both SWCNTs. However,  as illustrated in Fig. \ref{DOSCNT}, the two types of SWCNT have distinctively different densities of states, and hence distinctively different electronic structures.

 As expected for capped SWCNTs, the DOS reveals an electronic structure still dominated by $\pi$ states, albeit without the symmetry and characteristic fingerprints around the Fermi level that occur for the corresponding ``infinite" (6,6) and (10,0) SWCNTs \cite{Dresselhaus}, as a result of spatial confinement effects. Additionally, we display in Figure \ref{DOSCNT} the DOS of these systems for different values of applied macroscopic electrostatic field, i.e., 3, 5, 7, and 10 V/$\mu$m (low-field regime) and 1, 1.54, and 2 V/nm (a higher-field regime that is closer to the known field values at which FE occurs - which typically lie between 2 and 5 V/nm for an emitter with work function near 4.5 eV).
 
 For the low-field regime, the perturbation in the DOS is negligible for both structures, whereas for the higher-field regime the largest changes in the pattern are noticed around the Fermi level for the (6,6) SWCNT. In particular, for this latter case, the electronic states approach the Fermi level for higher electrostatic field values \cite{Edgcombe2019}, while for the (10,0) SWCNT the features near the Fermi level are preserved even for higher macroscopic electrostatic fields.

\begin{figure}[H]
\centering
\includegraphics [width=0.38\textwidth] {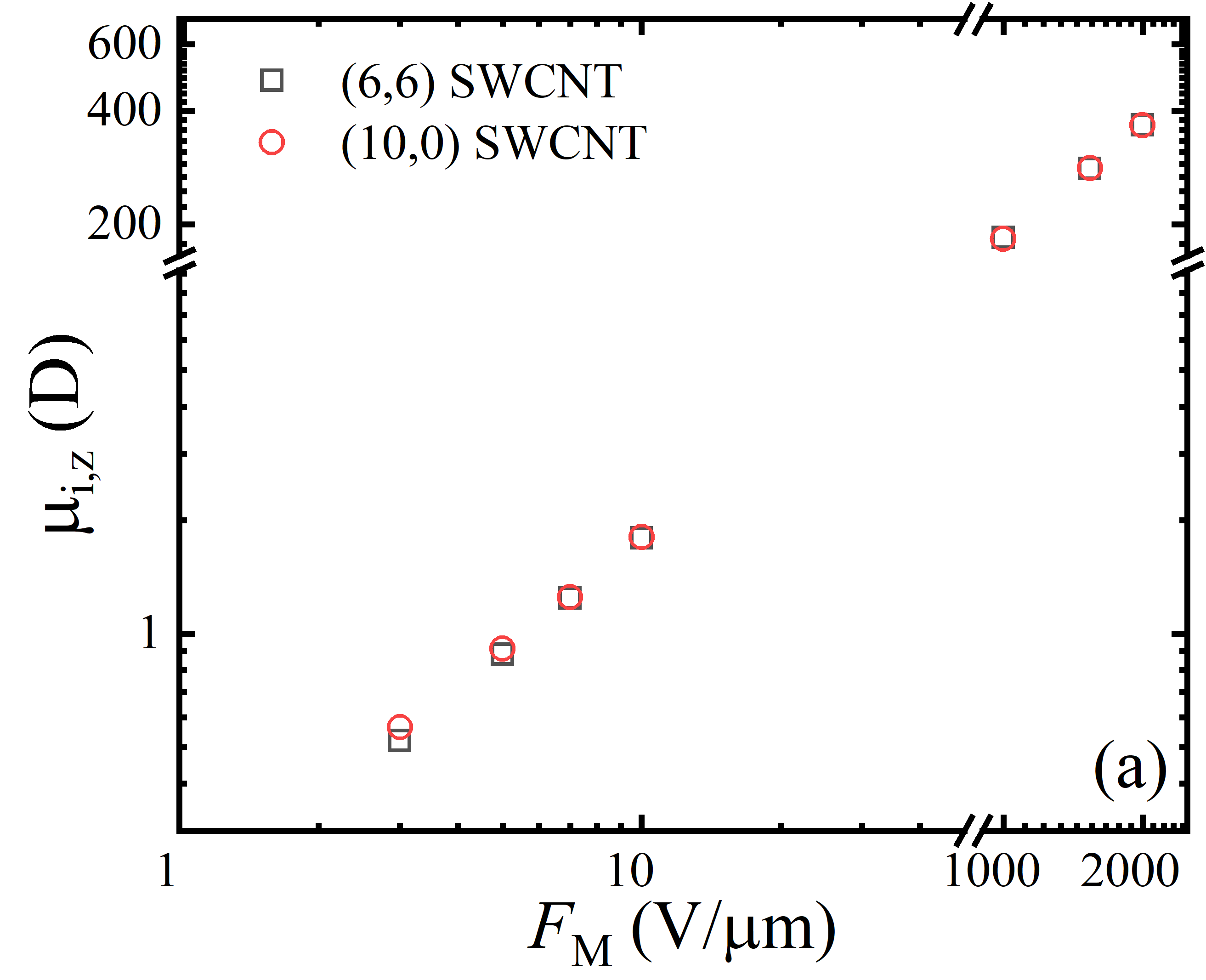} \\
\includegraphics [width=0.38\textwidth] {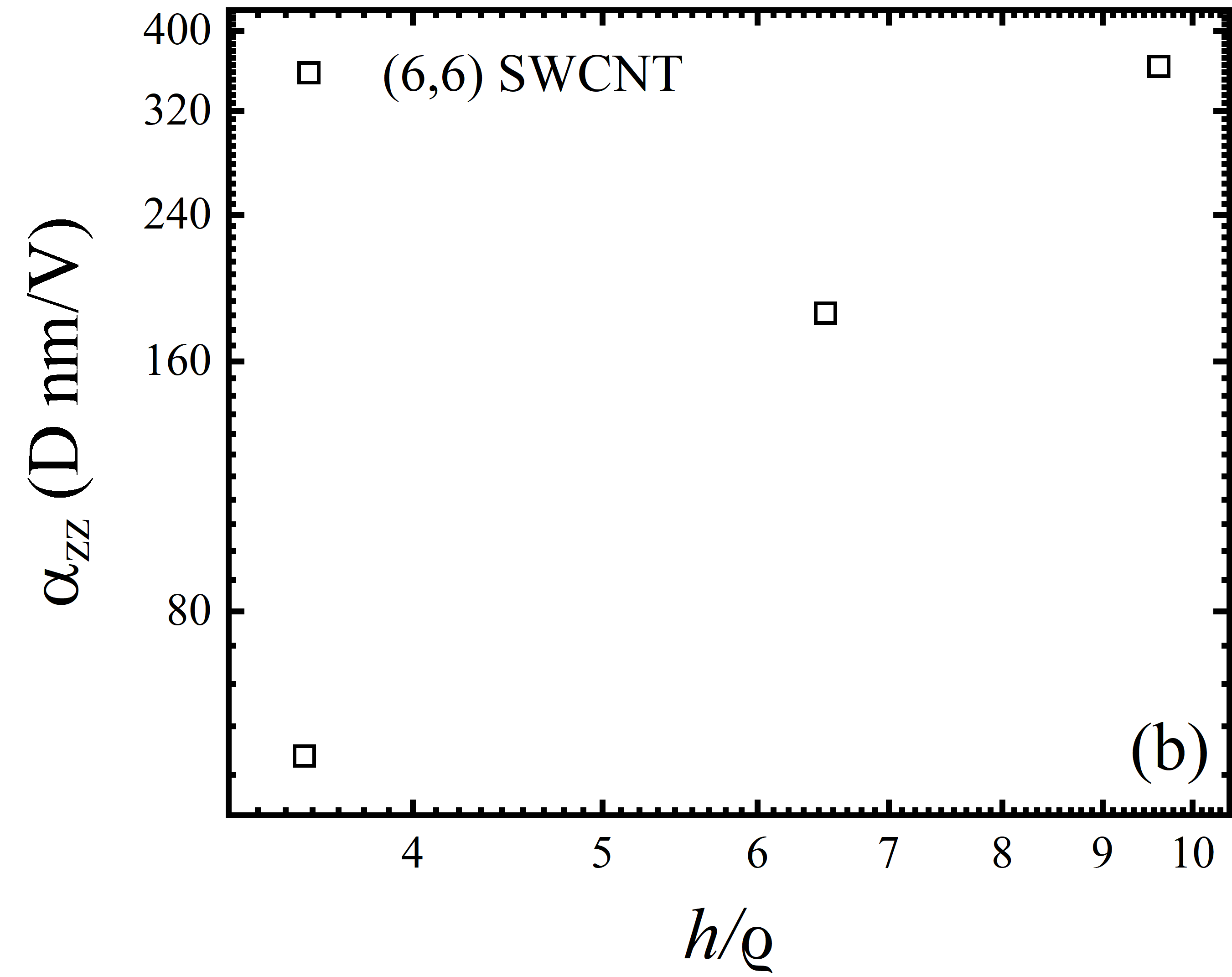}
\caption{(a) Calculated $z$-component of the induced Gaussian dipole moment $\mu_{{\rm{i}},z}$, for the (6,6) and (10,0) SWCNTs, measured in Debyes (D). (b) Hybrid longitudinal polarizability vs. aspect ratio, in log-log scale, for (6,6) SWCNTs with heights $h=1.48$, $2.73$ and $4.04$ nm and radii $\varrho = 0.42$nm.} \label{dipolem}
\end{figure}

In general, as with all polarizable systems, the effect of applying an electrostatic field to a SWCNT is to cause an increase in its system dipole moment, and this in turn is linked with local-field change in space near the negatively charged apex of the SWCNT. When the magnitude of the local field gets high enough, then field electron emission occurs. In Fig. \ref{dipolem} (a) we display, for both SWCNT types, the field dependence (on $F_{\rm{M}}$) of the longitudinal ($z$) component $\mu_{{\rm{i}},z}$ of the induced system dipole moment.

In the field regime used, a linear electric response is expected. In this case, we can write $\mu_{{\rm{i}},z} = \alpha_{zz} F_{\rm{M}}$, where $\alpha_{zz}$ is the longitudinal system polarizability \cite{JPCC2005} . From both the low and higher-field regimes and from both SWCNT types, we obtain the same linear function with respect to the applied macroscopic field.

This indicates that these structures exhibit the same longitudinal polarizability, regardless of their chiral indexes. The hybrid polarizability derived from Fig. \ref{dipolem} is about 182 D nm V$^{-1}$.  This is equivalent to a modern ISQ polarizability ($\alpha_{zz}$) of 6.07 $\times$10$^{-37}$ J m$^{2}$ V$^{-2}$ in SI units, or 3.31 $\times$10$^{4}$ eV nm$^{2}$ V$^{-2}$ in FE customary units.

Further, since both SWCNT structures have similar lengths and radii, this result leads the two systems to generate macroscopically similar electrostatic potential distributions. Thus, it is to be expected that the two types of SWCNT will generate similar values for characteristic FEFs. In Fig. \ref{dipolem} (b), we show the behavior of hybrid polarizability vs. $h/\varrho$ for (6,6) SWCNTs with heights $h=1.48$, $2.73$ and $4.04$ nm and radii $\varrho = 0.42$ nm. Our results suggest that polarizability increases when when $h/\varrho$ increases.

\subsection{Characteristic local induced field enhancement factor (LIFEF)}
\label{CFEF}
In classical-conductor theory, a parallel-planar-plate capacitor-like geometry is often discussed, where a HCP-model post stands on the plate designated as the emitter plate. The field enhancement at the post apex is caused by the response of the emitter's charge distribution to the application of the macroscopic field $F_{\rm{M}}$, in accordance with the rules of electron thermodynamics \cite{ForbesES}, which require that in static electrical equilibrium the Fermi level must be constant throughout the emitter. This results in a so-called classically induced charge distribution, and a related field distribution. In this system geometry, if the counter-electrode is sufficiently distant from the post apex, the emitter's characteristic FEF $\gamma_{\rm{C}}$ (which is usually taken in modelling as the apex FEF) is expected to depend only on the geometry of the post. As noted earlier, there are many theoretical treatments that aim to predict the value of $\gamma_{\rm{C}}$ as a function of the post height $h$ and radius $\varrho$. Classical treatments of this kind can be self-consistent even for very small values of $\varrho$. 

By contrast, in the quantum mechanics of CNTs, it was recognized many years ago \cite{APL2004china} that defining and predicting characteristic FEF values for SWCNTs would be a difficult enterprise, since CNTs are nanosystems in which details of the atomic structure are important.

However, the development in recent years of increasingly accurate DFT methods to treat thousands of atoms, has re-opened the issue of how to carry out detailed quantum-mechanical (QM) investigations into FE from SWCNTs. In investigations of this kind, there is still a significant procedural problem if one wants to simulate the true conditions used by experimentalists. The local field values at which FE occurs are typically around the range 2 to 5 V/nm. However, for various practical reasons, but especially to avoid the use of very high voltages that might lead to electrical breakdown effects, experimentalists prefer to use relatively low values of macroscopic field. This is turn requires CNTs with high apex FEFs, high values of the aspect ratio $h/\varrho$, and long nanotubes that contain very many atoms.

Early calculations (e.g., \cite{LiPRL,PRBLi2005,JAPForbes}) used QM/MM techniques of limited physical accuracy, but used CNTs that contained of order 60 000 atoms and generated physically realistic FEF values. Carrying out first-principles DFT calculations on a system of this size is operationally impossible at present, and likely to remain so for some considerable time. At present, DFT calculations can be carried out only on finite CNTs of very limited total length.   

One then has two alternatives. Investigate systems where macroscopic field values are realistic, but local apex field values are not.  Or investigate systems where the local apex field values are realistic but the macroscopic field values are much higher than those used in practice. In our work we initially chose the first of these options, but are now also exploring the second. We have been able to usefully investigate some of the physical issues relating to FEFs.

\begin{figure}[h!b]
\includegraphics [scale=0.5] {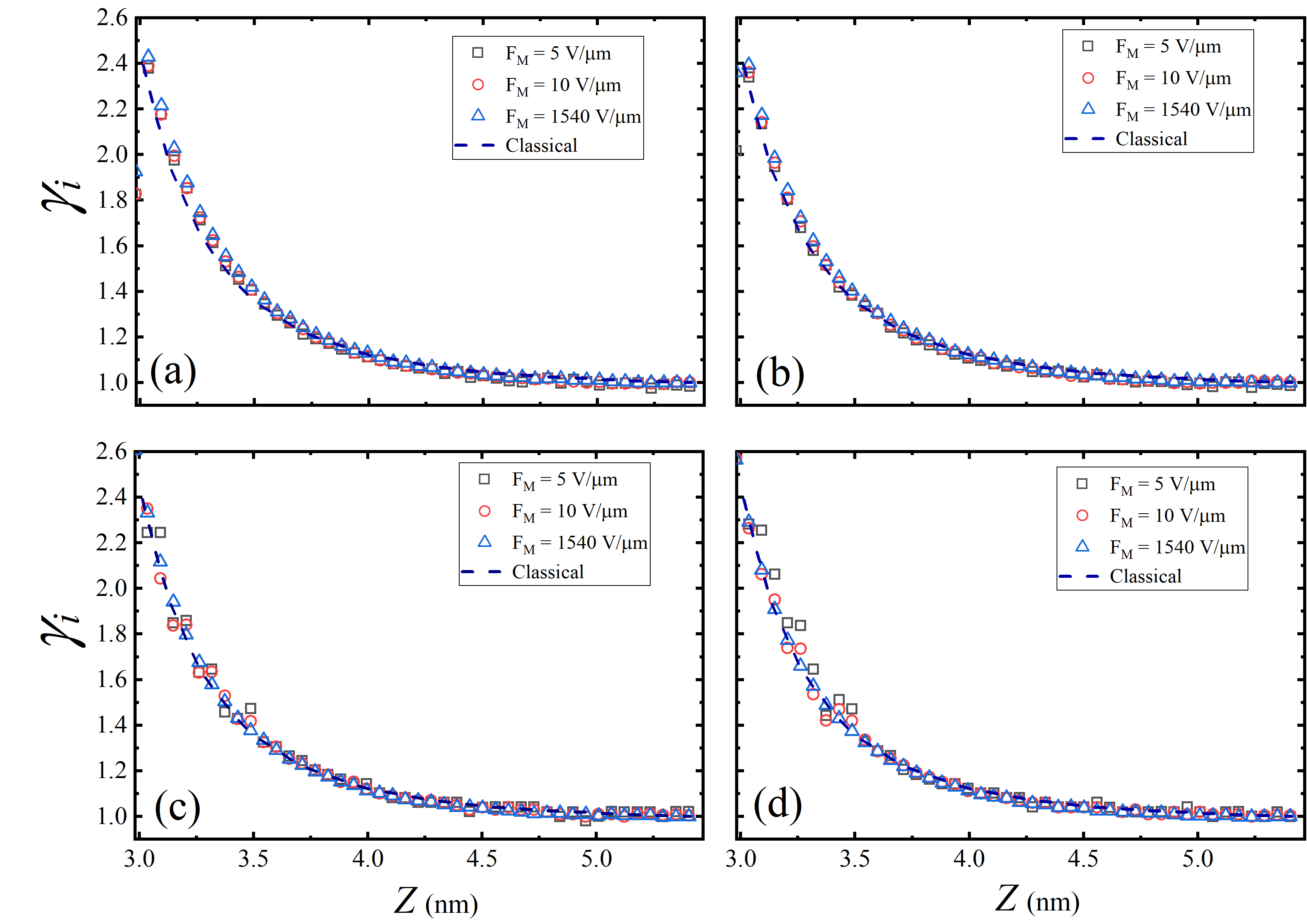}
\caption{Calculated LIFEF values as a function of distance $Z$ from the SWCNT mid-plane: (a), (b) for the (6,6) and (c), (d) for the (10,0) SWCNT, for $F_{\rm{M}}$-values as shown. Classical-conductor results for the HCP model are also shown. LIFEF variations along the CNT axis are shown in (a) and (c); LIFEF variations along a line parallel to the CNT axis, through one of the topmost atoms, are shown in (b) and (d).} \label{LIFEFCNTs}
\end{figure}
We are working with relatively small length CNTs (some hundreds of atoms) having sub-nanometric radii. An advantage is that we can investigate these nanosystems using a reliable first-principles methodology, namely DFT techniques. We have shown that the local induced FEF (LIFEF) remains constant when $F_{\rm{M}}$ changes from few V/$\mu$m to few V/nm. Figure \ref{LIFEFCNTs} shows the results for our calculated LIFEFs for the (6,6) and (10,0) floating SWCNTs, taken along the CNT axis and along a line, parallel to the axis, through one of the topmost atoms. In each case, results are shown for $F_{\rm{M}}$ $\in$ [3, 1540] V/$\mu$m and for $Z-Z_{\rm{a}} \geq 0.28$ nm. Here, $Z$ is distance measured along the CNT axis, from the ``mid-plane" that divides the CNT into positive and negative halves, and $Z_{\rm{a}}$ is the $Z$-value that corresponds to the average position of the six atoms of the topmost SWCNT ring. The distance 0.28 nm is the value of $Z_{\rm{R}}-Z_{\rm{a}}$, where $Z_{\rm{R}}$ is the position of a peak in the variation of the LIFEF with distance, and is the smallest $Z$-value for which the LIFEF can usefully be defined.

This position $Z_{\rm{R}}$ has been found to be independent of the macroscopic field and has been called the \textit{LIFEF reference point}. The LIFEF value there can be called the \textit{reference LIFEF} and regarded as characteristic of the CNT in question (though it remains to be established whether it coincides with the characteristic-FEF value that would be derived from a FN plot of the emission from a SWCNT with this geometry). 

As shown in Fig. \ref{LIFEFCNTs}, the LIFEF curves for different values of $F_{\rm{M}}$ clearly all collapse onto a single curve, thereby indicating that (at any given point along the line chosen) the LIFEF-value is independent of $F_{\rm{M}}$. FEF values predicted by the HCP classical-conductor model, using height and radius values equivalent to those characterising the relevant CNT, and finite element techniques, are also shown in the figure; these classical-conductor values coincide well with the LIFEF values obtained from the DFT calculations.

In so far as experimental results for CNTs are currently usually explained by using theoretical models (for PE variations) derived from classical-conductor models, the above theoretical results also appear compatible with the experimental result that straight-line FN plots are observed in the best experimental work on CNTs. (This FN-plot linearity implies that there exists a characteristic FEF that is constant, independent of $F_{\rm{M}}$.)   

The consistency between the classical results and the DFT-obtained induced-FEF results can be interpreted as a consequence of static linear response theory. As previously discussed (see Fig. \ref{dipolem}), the $z$-component of the system dipole moment, $\mu_{{\rm{i}},z}$, is found to change linearly with $F_{\rm{M}}$ in both low-field and higher-field ranges. For zero $F_{\rm{M}}$, the CNT system dipole moment is zero, because of the effective symmetry about the mid-plane. Thus, for non-zero $F_{\rm{M}}$, the induced dipole moment (which is the \textit{change} in system dipole moment) is equal to the real dipole moment as calculated using DFT. Thus, Fig. \ref{dipolem} can be taken as representing the behavior of the induced system dipole moment.

Then, using Eq. (\ref{LIFEF}) and assuming a linear polarization regime, the reference LIFEF $\gamma_{{\rm{i}},{\rm{R}}}$ can be written in the form

\begin{equation}
\gamma_{{\rm{i}},{\rm{R}}} = \left(\frac{F_{\rm{i},\rm{R}}}{\mu_{{\rm{i}},z}}\right) \alpha_{zz},
\label{response1}
\end{equation}
where $F_{\rm{i,R}}$ is the local induced field at the reference position ``R", and $\alpha_{zz}$ (as before) is the system longitudinal polarizability.

\begin{figure}[h!]
\includegraphics [scale=0.3] {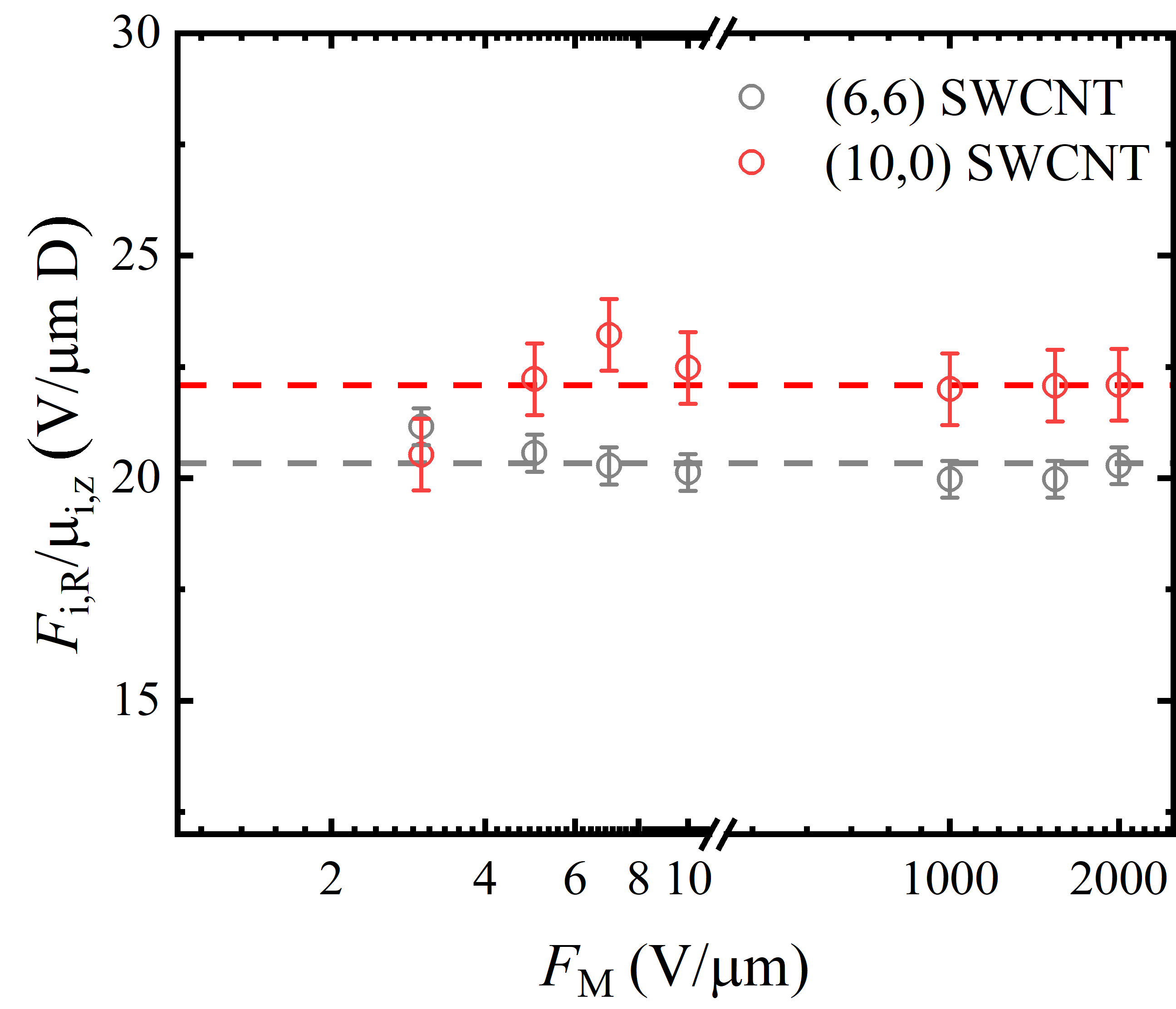}
\caption{The $(F_{\rm{i,R}}/\mu_{{\rm{i}},z})$ ratio calculated as a function of $F_{\rm{M}}$, for relevant ranges of $F_{\rm{M}}$, for the (6,6) and (10,0) SWCNTs. Error bars are also shown. Horizontal dashed lines correspond to average values of $(F_{\rm{i,R}}/\mu_{{\rm{i}},z})$.} \label{constancy}
\end{figure}

Since LIFEF values are independent of $F_{\rm{M}}$, the  bracketed term $(F_{\rm{i,R}}/\mu_{{\rm{i}},z})$ in Eq.(\ref{response1}) should also be constant. In Fig. \ref{constancy}, our results clearly corroborate this conclusion for the (6,6) and (10,0) floating SWCNTs. Moreover, this ratio $({F_{\rm{i,R}}}/{\mu_{{\rm{i}},z}})$ is very similar for these two types of SWCNT. Since the longitudinal system polarizability for these structures appears to be the same, the similarity of the reference-LIFEFs reported in Fig. \ref{LIFEFCNTs} for the two SWCNT structures is to be expected. This analysis leads to the conclusion that the reference-LIFEF can be understood as dependent only on parameters related to the structural properties of the CNTs, namely (a) the ratio $({F_{\rm{i,R}}}/{\mu_{{\rm{i}},z}})$, and (b) the response of the material to the applied macroscopic field $F_{\rm{M}}$, as assessed by the longitudinal system polarizability $\alpha_{zz}$.

It is clear from this study that the behavior of the SWCNTs can be described as follows. The applied macroscopic electrostatic field creates a perturbation that acts on the SWCNT ground electronic state. This perturbation can be described by an ``external" electrostatic-potential term of the form  $V_{\rm{ext}}(\textbf{r})$. Applying this perturbation leads to a small change in the SWCNT electron density. Using the static linear response theory, the resulting induced electron density $\rho_{\rm{i}}(\mathbf{r},F_{\rm{M}})$ is adequately given by the first term in the expansion

\begin{equation}
\begin{split}
\rho_{\rm{i}}(\mathbf{r},F_{\rm{M}}) = \int\chi_{\rm{e}}(\textbf{r},\textbf{r}')\Delta V_{\rm{ext}}(\textbf{r}'){\rm{d}}\textbf{r}'+ O[(\Delta V_{\rm{ext}})^2].
\label{response}
\end{split}
\end{equation}
In Eq. (\ref{response}), the function $\chi_{\rm{e}}(\textbf{r},\textbf{r}')$ is called the static density response function. As well-known, linear response theory is concerned with its properties and functional form. It is known from this theory \cite{parr1994density} that the density response function is specific to the system of interest and a functional of the ground-state electron density distribution, as it exists in the absence of any applied macroscopic field. 

In our case, $\chi_{\rm{e}}(\textbf{r},\textbf{r}')$ is connected with Hohenberg-Kohn-Sham density functional theory \cite{Onida}. Further, the reference LIFEF $\gamma_{\rm{i,R}}$ is defined in terms of the induced electron density. If it is valid to use perturbation theory based on the zero-$F_{\rm{M}}$ ground-electronic-state wave-functions (as normally assumed in linear response theory), then the value of the reference LIFEF will depend on the response function $\chi_{\rm{e}}$. Thus, this reference LIFEF $\gamma_{\rm{i,R}}$ value, identified as proposed here, can presumably be seen as a parameter that is a physical consequence of how the charge-density distribution responds to $F_{\rm{M}}$ in a linear regime, and thus as a parameter that characterises the SWCNT of interest. 

This viewpoint is reinforced by the observation that our results are fairly similar for the two distinct SWCNT structures and for the two observation-lines examined (one being the CNT axis, the other a line through a topmost surface atom, parallel to the axis). As already noted, the calculated system polarizabilities for the (6,6) and (10,0) floating SWCNTs are very similar, as shown in Fig. \ref{dipolem}.

\begin{figure}[h!]
\includegraphics [scale=0.5] {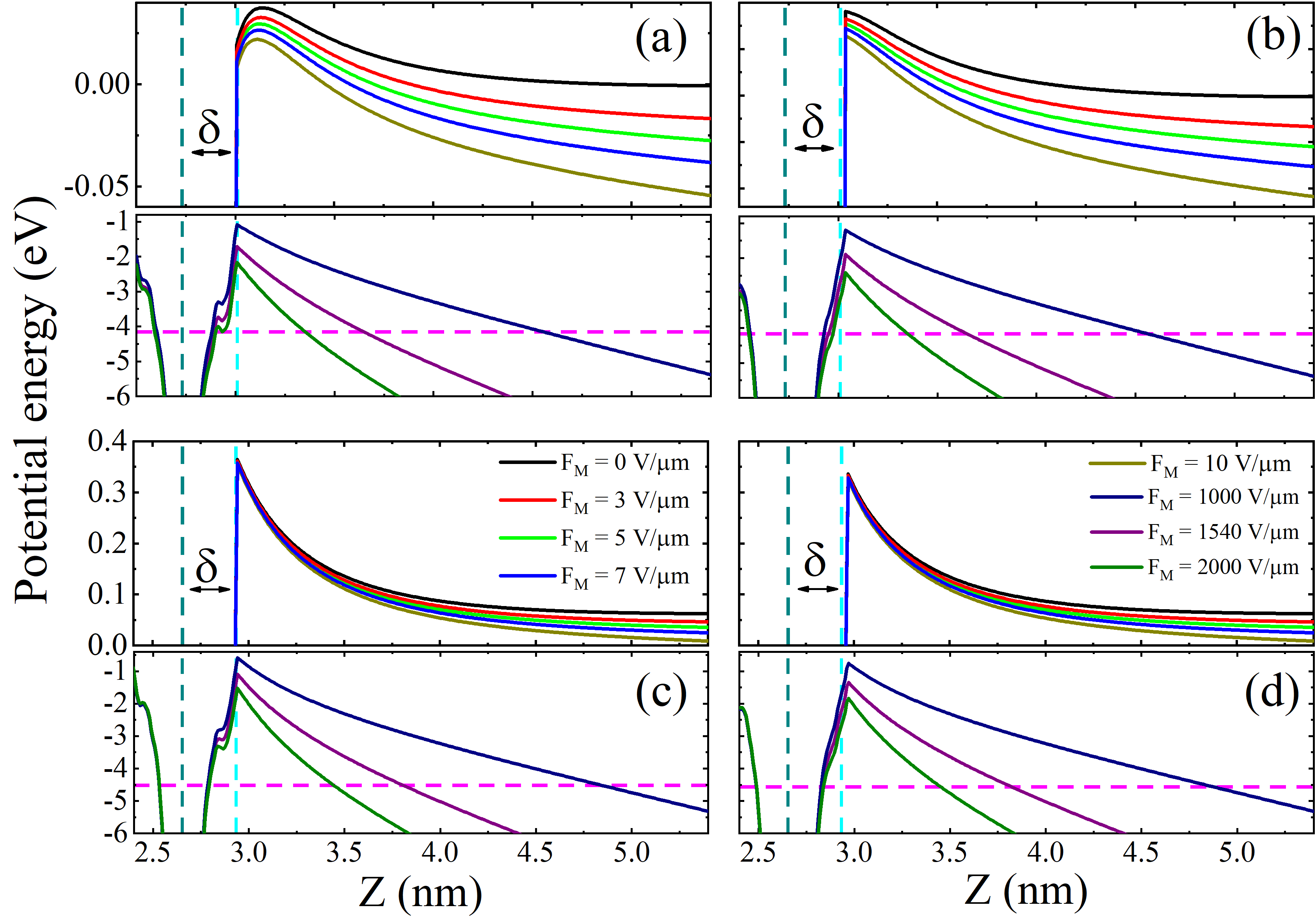}
\caption{Potential-energy (PE) barriers change as a function of macroscopic field $F_{\rm{M}}$: (a), (b) for the (6,6) SWCNT, and (c), (d) for the (10,0) SWCNT. Left panels show the PE variation along the axis, right panels the PE variation along a parallel line through one of the topmost atom. Top panels represent the low-$F_{\rm{M}}$ regime and bottom panels the higher-$F_{\rm{M}}$ regime. Left and right vertical dashed lines show the average position $Z_{\rm{a}}$ of the hexagonal-ring topmost atoms, and the position $Z_{\rm{R}}$ of the LIFEF reference point. The quantity $\delta = Z_{\rm{R}} - Z_{\rm{a}}$ is also shown. Horizontal dashed lines indicate the Fermi level calculated at the DFT level PBE/DZP.} \label{Pot}
\end{figure}

Finally, we stress again that the LIFEFs appear to have the same general properties as the local FEFs derived from a classical-conductor HCP model, as was shown in Fig.\ref{LIFEFCNTs}. These classical FEFs were calculated using a finite-element methodology, in the form of the COMSOL code \cite{JVSTB2019}. The same rectangular simulation boxes with a square base, and the same dimensions for the simulation box and for the HCP model, were used as in DFT simulations of the SWCNTs, described previously.

\subsection{Potential energy barriers}

Electron tunneling through the surface potential-energy (PE) barrier, and hence the related emission current density, is highly sensitive to the local electrostatic-field value $F_{\rm{L}}$ that defines the tunnelling barrier. Since, for any given surface location across the surface, $F_{\rm{L}} = \gamma_{\rm{L}} F_{\rm{M}}$, where $\gamma_{\rm{L}}$ is the relevant local FEF-value, local emission current densities are very sensitive to the applied macroscopic field $F_{\rm{M}}$.

What this means is that the barrier height and width (which depend primarily on the barrier zero-field height and on the local-field value $F_{\rm{L}}$), become---at any given surface location---sensitive functions of macroscopic field $F_{\rm{M}}$. In Figure \ref{Pot}, the results for some ``real" PE barriers are shown for $F_{\rm{M}}$ $\in$ [3, 2000] V/$\mu$m. Results are shown for both of the SWCNT structures examined and for the two ``lines of observation" used previously. Detailed investigation of these results, and exploration of how they relate to the equivalent results for ``induced" PE barriers, are still ``work in progress". However, these diagrams have some interesting features that it seems worth pointing out at this stage.

For the ``higher-field" $F_{\rm{M}}$ range (1 to 2 V/nm), which yields local fields close to those used in practical field electron emission, the results in all cases are fairly similar. But for the lower-field range there are significant differences, particularly between the two SWCNT structures. In the diagrams, the zero of energy corresponds to the ``distant vacuum level", i.e., the potential energy of an electron at a location that is far away from the floating CNT, in comparison with the CNT dimensions, when the applied macroscopic field is zero. For the (6,6) floating SWCNT this distant vacuum level is 4.16 eV above the SWCNT Fermi level; for the (10,0) floating SWCNT, the distant vacuum level is 4.56 eV above the SWCNT Fermi level. For the $F_{\rm{M}}=0$ case, for both SWCNT structures, the energy level of the top of the local barrier is positive, i.e., it is above the distant vacuum level. For the (6,6) structure the local-barrier-top level is only slightly positive (by less than 0.05 eV), but for the (10.0) structure the local-barrier-top level is positive by over 0.3 eV. 

There are also differences between the heights of the local barriers above the related Fermi levels. For example, in the case of the PE variations along a line through one of the topmost SWCNT atoms, for $F_{\rm{M}}$ =1.54 V/nm, the barrier height for the (6,6) SWCNT was found to be 2.17 eV, compared with the value 3.21 eV for the (10,0) SWCNT. These results appear to indicate that there must be chemically induced charge transfers taking place near the apexes of the SWCNTs, and that these charge-transfers result in a system of \textit{patch fields} surrounding the CNT apex. Further (not surprisingly), the exact nature of these charge transfers would appear to depend on the SWCNT structure. This thinking is consistent with previous thinking concerning charge behaviour near the emitter apex, as found when modelling CNTs with lengths of the order of micrometers \cite{JAPForbes}.

\section{Conclusions}
\label{conclusions}	

In summary, we have used density functional theory and DFT-based techniques to determine ``real" and ``induced" charge and field distributions for small floating SWCNTs capped at both ends, and to determine related local induced field enhancement factors (LIFEFs). Calculations were performed for (6,6) and (10,0) SWCNT structures, taking into account the changes in electron densities induced by an applied macroscopic electrostatic field, $F_{\rm{M}}$. We have found that, at all points in space outside the SWCNT apex, the calculated LIFEF values are independent of macroscopic field (as is the case for classical-conductor models of SWCNTs). This allows the definition of a ``reference LIFEF value" that is characteristic of the SWCNT.

Interestingly, in both the low-$F_{\rm{M}}$ (a few V/$\mu$m) and higher-$F_{\rm{M}}$ (a few V/nm) regimes, the $z$-component of the induced system dipole moment, as calculated for both types of SWCNTs, is effectively the same linear function of $F_{\rm{M}}$, indicating that these structures possess similar longitudinal system polarizabilities. As a consequence, these systems also exhibit similar $F_{\rm{M}}$-independent LIFEFs. Our results suggest that a defined ``reference LIFEF" could be understood as dependent only on the structural parameters of the SWCNT and its response to the applied macroscopic field $F_{\rm{M}}$, and that a close connection with linear response theory probably exists. We have also found evidence to confirm that chemically-induced charge-transfer effects occur at the apexes of capped SWCNTs, and lead to patch-field systems at the apex. Work continues in order to develop a better understanding of how best (if possible) to create links between real field electron emission from SWCNTs, and DFT models thereof, and conventional FE theory.

\begin{acknowledgement}

This study was partially funded by the Brazilian agencies: Coordena\c{c}\~{a}o de Aperfei\c{c}oamento de Pessoal de N\'{i}vel Superior (CAPES), Finance Code 001, Conselho Nacional de Desenvolvimento Cient\'{i}fico e Tecnol\'{o}gico (CNPq), and Funda\c{c}\~{a}o de Amparo \`{a} Pesquisa do Estado da Bahia (FAPESB). RR also thanks INCT-FCx.
\end{acknowledgement}

\vspace{0.5cm}

\section{Author Information}
\subsection{Corresponding Author}
*E-mail: thiagoaa@ufba.br; phone: +55 71 32836667; fax: +55 71 32836606. \subsection{ORCID}
Caio P. de Castro: 0000-0001-5316-7515;
Thiago A. de Assis: 0000-0003-2332-1096;
Roberto Rivelino: 0000-0003-2679-1640;
Fernando de B. Mota: 0000-0001-9571-8549;
Caio M. C. de Castilho: 0000-0003-3788-4412;
Richard G. Forbes: 0000-0002-8621-3298.

Notes: The authors declare no competing financial interest.


\providecommand{\latin}[1]{#1}
\makeatletter
\providecommand{\doi}
  {\begingroup\let\do\@makeother\dospecials
  \catcode`\{=1 \catcode`\}=2 \doi@aux}
\providecommand{\doi@aux}[1]{\endgroup\texttt{#1}}
\makeatother
\providecommand*\mcitethebibliography{\thebibliography}
\csname @ifundefined\endcsname{endmcitethebibliography}
  {\let\endmcitethebibliography\endthebibliography}{}


\begin{tocentry}

\begin{figure}[H]
	\centering    %
    \vspace{-0.3cm}
     \includegraphics[width=7.5cm,height=3.7cm]{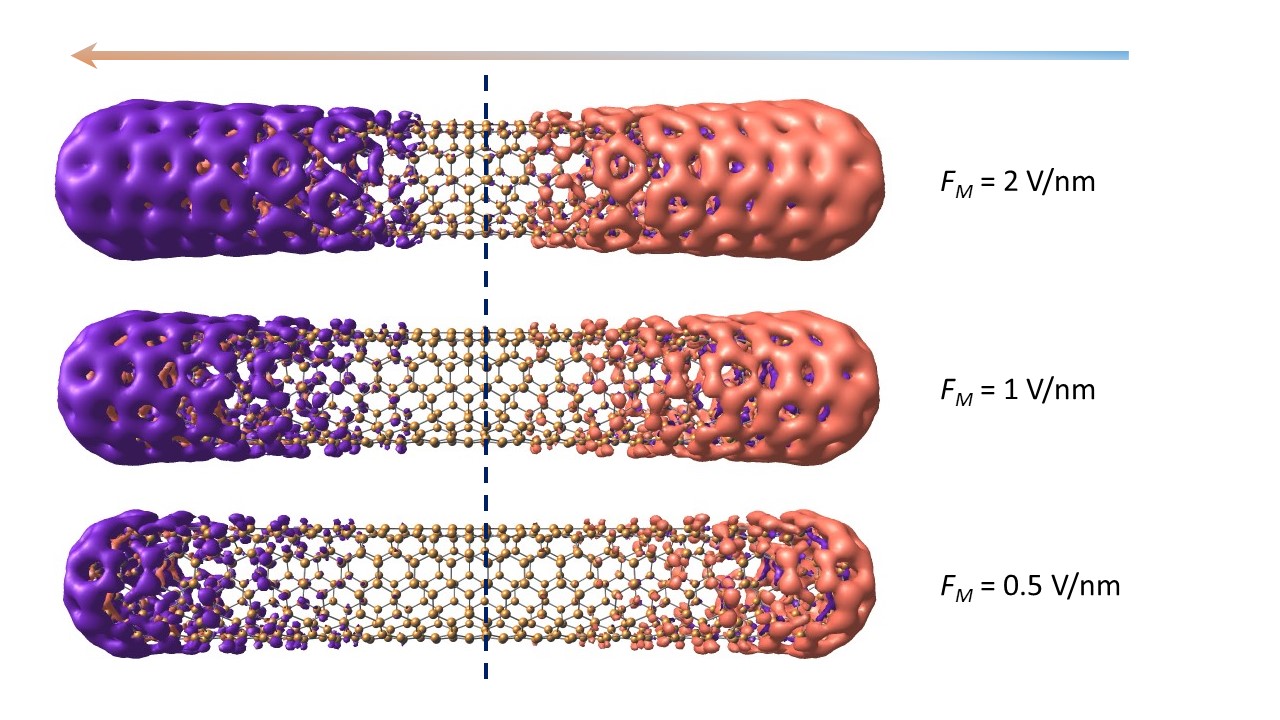}
\label{Arte}
\end{figure}
%





\end{tocentry}


\begin{mcitethebibliography}{40}
\providecommand*\natexlab[1]{#1}
\providecommand*\mciteSetBstSublistMode[1]{}
\providecommand*\mciteSetBstMaxWidthForm[2]{}
\providecommand*\mciteBstWouldAddEndPuncttrue
  {\def\EndOfBibitem{\unskip.}}
\providecommand*\mciteBstWouldAddEndPunctfalse
  {\let\EndOfBibitem\relax}
\providecommand*\mciteSetBstMidEndSepPunct[3]{}
\providecommand*\mciteSetBstSublistLabelBeginEnd[3]{}
\providecommand*\EndOfBibitem{}
\mciteSetBstSublistMode{f}
\mciteSetBstMaxWidthForm{subitem}{(\alph{mcitesubitemcount})}
\mciteSetBstSublistLabelBeginEnd
  {\mcitemaxwidthsubitemform\space}
  {\relax}
  {\relax}

\bibitem[Cole \latin{et~al.}(2015)Cole, Mann, Teo, and Milne]{Cole2015chapter}
Cole,~M.~T.; Mann,~M.; Teo,~K.~B.; Milne,~W.~I. In \emph{Emerging
  Nanotechnologies for Manufacturing (Second Edition)}, second edition ed.;
  Ahmed,~W., , Jackson,~M.~J., Eds.; Micro and Nano Technologies; William
  Andrew Publishing: Boston, 2015; pp 125 -- 186\relax
\mciteBstWouldAddEndPuncttrue
\mciteSetBstMidEndSepPunct{\mcitedefaultmidpunct}
{\mcitedefaultendpunct}{\mcitedefaultseppunct}\relax
\EndOfBibitem
\bibitem[Esat \latin{et~al.}(2018)Esat, Esat, Friedrich, Tautz, and
  Temirov]{nature2018}
Esat,~T.; Esat,~T.; Friedrich,~N.; Tautz,~F.~S.; Temirov,~R. A standing
  molecule as a single-electron field emitter. \emph{Nature} \textbf{2018},
  \emph{558}, 573--578\relax
\mciteBstWouldAddEndPuncttrue
\mciteSetBstMidEndSepPunct{\mcitedefaultmidpunct}
{\mcitedefaultendpunct}{\mcitedefaultseppunct}\relax
\EndOfBibitem
\bibitem[Buldum and Lu(2003)Buldum, and Lu]{Buldum}
Buldum,~A.; Lu,~J.~P. Electron Field Emission Properties of Closed Carbon
  Nanotubes. \emph{Phys. Rev. Lett.} \textbf{2003}, \emph{91}, 236801\relax
\mciteBstWouldAddEndPuncttrue
\mciteSetBstMidEndSepPunct{\mcitedefaultmidpunct}
{\mcitedefaultendpunct}{\mcitedefaultseppunct}\relax
\EndOfBibitem
\bibitem[Vincent \latin{et~al.}(2011)Vincent, Poncharal, Barois, Perisanu,
  Gouttenoire, Frachon, Lazarus, de~Langre, Minoux, Charles, Ziaei, Guillot,
  Choueib, Ayari, and Purcell]{Vicent}
Vincent,~P.; Poncharal,~P.; Barois,~T.; Perisanu,~S.; Gouttenoire,~V.;
  Frachon,~H.; Lazarus,~A.; de~Langre,~E.; Minoux,~E.; Charles,~M.; Ziaei,~A.;
  Guillot,~D.; Choueib,~M.; Ayari,~A.; Purcell,~S.~T. Performance of
  field-emitting resonating carbon nanotubes as radio-frequency demodulators.
  \emph{Phys. Rev. B} \textbf{2011}, \emph{83}, 155446\relax
\mciteBstWouldAddEndPuncttrue
\mciteSetBstMidEndSepPunct{\mcitedefaultmidpunct}
{\mcitedefaultendpunct}{\mcitedefaultseppunct}\relax
\EndOfBibitem
\bibitem[Pascale-Hamri \latin{et~al.}(2014)Pascale-Hamri, Perisanu, Derouet,
  Journet, Vincent, Ayari, and Purcell]{Pascale}
Pascale-Hamri,~A.; Perisanu,~S.; Derouet,~A.; Journet,~C.; Vincent,~P.;
  Ayari,~A.; Purcell,~S.~T. Ultrashort Single-Wall Carbon Nanotubes Reveal
  Field-Emission Coulomb Blockade and Highest Electron-Source Brightness.
  \emph{Phys. Rev. Lett.} \textbf{2014}, \emph{112}, 126805\relax
\mciteBstWouldAddEndPuncttrue
\mciteSetBstMidEndSepPunct{\mcitedefaultmidpunct}
{\mcitedefaultendpunct}{\mcitedefaultseppunct}\relax
\EndOfBibitem
\bibitem[Cole \latin{et~al.}(2016)Cole, Parmee, and Milne]{Nanotech2016}
Cole,~M.~T.; Parmee,~R.~J.; Milne,~W.~I. Nanomaterial-based x-ray sources.
  \emph{Nanotechnology} \textbf{2016}, \emph{27}, 082501\relax
\mciteBstWouldAddEndPuncttrue
\mciteSetBstMidEndSepPunct{\mcitedefaultmidpunct}
{\mcitedefaultendpunct}{\mcitedefaultseppunct}\relax
\EndOfBibitem
\bibitem[Min \latin{et~al.}(2019)Min, Kim, Moon, Han, Yum, and Lee]{Hyegi}
Min,~H.; Kim,~Y.-T.; Moon,~S.~M.; Han,~J.-H.; Yum,~K.; Lee,~C.~Y. High-Yield
  Fabrication, Activation, and Characterization of Carbon Nanotube Ion Channels
  by Repeated Voltage-Ramping of Membrane-Capillary Assembly. \emph{Advanced
  Functional Materials} \textbf{2019}, \emph{29}, 1900421\relax
\mciteBstWouldAddEndPuncttrue
\mciteSetBstMidEndSepPunct{\mcitedefaultmidpunct}
{\mcitedefaultendpunct}{\mcitedefaultseppunct}\relax
\EndOfBibitem
\bibitem[Nguyen \latin{et~al.}(2019)Nguyen, Wallum, Nguyen, Nguyen, Lyding, and
  Gruebele]{Nguyen}
Nguyen,~D.; Wallum,~A.; Nguyen,~H.~A.; Nguyen,~N.~T.; Lyding,~J.~W.;
  Gruebele,~M. Imaging of Carbon Nanotube Electronic States Polarized by the
  Field of an Excited Quantum Dot. \emph{ACS Nano} \textbf{2019}, \emph{13},
  1012\relax
\mciteBstWouldAddEndPuncttrue
\mciteSetBstMidEndSepPunct{\mcitedefaultmidpunct}
{\mcitedefaultendpunct}{\mcitedefaultseppunct}\relax
\EndOfBibitem
\bibitem[Tans \latin{et~al.}(1998)Tans, Verschueren, and Dekker]{Tans}
Tans,~S.~J.; Verschueren,~A. R.~M.; Dekker,~C. Room-temperature transistor
  based on a single carbon nanotube. \emph{Nature} \textbf{1998}, \emph{393},
  49\relax
\mciteBstWouldAddEndPuncttrue
\mciteSetBstMidEndSepPunct{\mcitedefaultmidpunct}
{\mcitedefaultendpunct}{\mcitedefaultseppunct}\relax
\EndOfBibitem
\bibitem[Wong \latin{et~al.}(1998)Wong, Joselevich, Woolley, Cheung, and
  Lieber]{Wong}
Wong,~S.~S.; Joselevich,~E.; Woolley,~A.~T.; Cheung,~C.~L.; Lieber,~C.~M.
  Covalently functionalized nanotubes as nanometre- sized probes in chemistry
  and biology. \emph{Nature} \textbf{1998}, \emph{394}, 52\relax
\mciteBstWouldAddEndPuncttrue
\mciteSetBstMidEndSepPunct{\mcitedefaultmidpunct}
{\mcitedefaultendpunct}{\mcitedefaultseppunct}\relax
\EndOfBibitem
\bibitem[Bonard \latin{et~al.}(2002)Bonard, Dean, Coll, and
  Klinke]{BonardPRL2002}
Bonard,~J.-M.; Dean,~K.~A.; Coll,~B.~F.; Klinke,~C. Field Emission of
  Individual Carbon Nanotubes in the Scanning Electron Microscope. \emph{Phys.
  Rev. Lett.} \textbf{2002}, \emph{89}, 197602\relax
\mciteBstWouldAddEndPuncttrue
\mciteSetBstMidEndSepPunct{\mcitedefaultmidpunct}
{\mcitedefaultendpunct}{\mcitedefaultseppunct}\relax
\EndOfBibitem
\bibitem[Dean and Chalamala(2000)Dean, and Chalamala]{Dean2000}
Dean,~K.~A.; Chalamala,~B.~R. Current saturation mechanisms in carbon nanotube
  field emitters. \emph{Applied Physics Letters} \textbf{2000}, \emph{76},
  375--377\relax
\mciteBstWouldAddEndPuncttrue
\mciteSetBstMidEndSepPunct{\mcitedefaultmidpunct}
{\mcitedefaultendpunct}{\mcitedefaultseppunct}\relax
\EndOfBibitem
\bibitem[Forbes(2013)]{Forbes2013}
Forbes,~R.~G. Development of a simple quantitative test for lack of field
  emission orthodoxy. \emph{Proceedings of the Royal Society of London A:
  Mathematical, Physical and Engineering Sciences} \textbf{2013}, \emph{469},
  20130271\relax
\mciteBstWouldAddEndPuncttrue
\mciteSetBstMidEndSepPunct{\mcitedefaultmidpunct}
{\mcitedefaultendpunct}{\mcitedefaultseppunct}\relax
\EndOfBibitem
\bibitem[Parr and Weitao(1994)Parr, and Weitao]{parr1994density}
Parr,~R.; Weitao,~Y. \emph{Density-Functional Theory of Atoms and Molecules};
  International Series of Monographs on Chemistry; Oxford University Press,
  1994\relax
\mciteBstWouldAddEndPuncttrue
\mciteSetBstMidEndSepPunct{\mcitedefaultmidpunct}
{\mcitedefaultendpunct}{\mcitedefaultseppunct}\relax
\EndOfBibitem
\bibitem[Forbes(1999)]{ForbesES}
Forbes,~R.~G. The electrical surface as centroid of the surface-induced charge.
  \emph{Ultramicroscopy} \textbf{1999}, \emph{79}, 25 -- 34\relax
\mciteBstWouldAddEndPuncttrue
\mciteSetBstMidEndSepPunct{\mcitedefaultmidpunct}
{\mcitedefaultendpunct}{\mcitedefaultseppunct}\relax
\EndOfBibitem
\bibitem[de~Castro \latin{et~al.}(2019)de~Castro, de~Assis, Rivelino,
  de~B.~Mota, de~Castilho, and Forbes]{JPCC2019}
de~Castro,~C.~P.; de~Assis,~T.~A.; Rivelino,~R.; de~B.~Mota,~F.;
  de~Castilho,~C. M.~C.; Forbes,~R.~G. Restoring Observed Classical Behavior of
  the Carbon Nanotube Field Emission Enhancement Factor from the Electronic
  Structure. \emph{The Journal of Physical Chemistry C} \textbf{2019},
  \emph{123}, 5144--5149\relax
\mciteBstWouldAddEndPuncttrue
\mciteSetBstMidEndSepPunct{\mcitedefaultmidpunct}
{\mcitedefaultendpunct}{\mcitedefaultseppunct}\relax
\EndOfBibitem
\bibitem[de~Castro \latin{et~al.}(2019)de~Castro, de~Assis, Rivelino, B.~Mota,
  de~Castilho, and Forbes]{deCastro2019jap}
de~Castro,~C.~P.; de~Assis,~T.~A.; Rivelino,~R.; B.~Mota,~F.~d.;
  de~Castilho,~C. M.~C.; Forbes,~R.~G. On the quantum mechanics of how an ideal
  carbon nanotube field emitter can exhibit a constant field enhancement
  factor. \emph{Journal of Applied Physics} \textbf{2019}, \emph{126},
  204302\relax
\mciteBstWouldAddEndPuncttrue
\mciteSetBstMidEndSepPunct{\mcitedefaultmidpunct}
{\mcitedefaultendpunct}{\mcitedefaultseppunct}\relax
\EndOfBibitem
\bibitem[Peng \latin{et~al.}(2005)Peng, Li, He, Deng, Xu, Zheng, and
  Chen]{PRBLi2005}
Peng,~J.; Li,~Z.; He,~C.; Deng,~S.; Xu,~N.; Zheng,~X.; Chen,~G. Quantum
  mechanical understanding of field dependence of the apex barrier of a
  single-wall carbon nanotube. \emph{Phys. Rev. B} \textbf{2005}, \emph{72},
  235106\relax
\mciteBstWouldAddEndPuncttrue
\mciteSetBstMidEndSepPunct{\mcitedefaultmidpunct}
{\mcitedefaultendpunct}{\mcitedefaultseppunct}\relax
\EndOfBibitem
\bibitem[Edgcombe and Valdr\`{e}(2001)Edgcombe, and Valdr\`{e}]{Edgcombe2001}
Edgcombe,~C.~J.; Valdr\`{e},~U. Microscopy and computational modelling to
  elucidate the enhancement factor for field electron emitters. \emph{Journal
  of Microscopy} \textbf{2001}, \emph{203}, 188--194\relax
\mciteBstWouldAddEndPuncttrue
\mciteSetBstMidEndSepPunct{\mcitedefaultmidpunct}
{\mcitedefaultendpunct}{\mcitedefaultseppunct}\relax
\EndOfBibitem
\bibitem[Forbes \latin{et~al.}(2003)Forbes, Edgcombe, and Valdr\`{e}]{Edgcombe}
Forbes,~R.~G.; Edgcombe,~C.; Valdr\`{e},~U. Some comments on models for field
  enhancement. \emph{Ultramicroscopy} \textbf{2003}, \emph{95}, 57 -- 65\relax
\mciteBstWouldAddEndPuncttrue
\mciteSetBstMidEndSepPunct{\mcitedefaultmidpunct}
{\mcitedefaultendpunct}{\mcitedefaultseppunct}\relax
\EndOfBibitem
\bibitem[Liang and Xu(2003)Liang, and Xu]{APL2000china}
Liang,~S.-D.; Xu,~N.~S. Chirality effect of single-wall carbon nanotubes on
  field emission. \emph{Applied Physics Letters} \textbf{2003}, \emph{83},
  1213--1215\relax
\mciteBstWouldAddEndPuncttrue
\mciteSetBstMidEndSepPunct{\mcitedefaultmidpunct}
{\mcitedefaultendpunct}{\mcitedefaultseppunct}\relax
\EndOfBibitem
\bibitem[Liang \latin{et~al.}(2004)Liang, Huang, Deng, and Xu]{APL2004china}
Liang,~S.-D.; Huang,~N.~Y.; Deng,~S.~Z.; Xu,~N.~S. Chiral and quantum size
  effects of single-wall carbon nanotubes on field emission. \emph{Applied
  Physics Letters} \textbf{2004}, \emph{85}, 813--815\relax
\mciteBstWouldAddEndPuncttrue
\mciteSetBstMidEndSepPunct{\mcitedefaultmidpunct}
{\mcitedefaultendpunct}{\mcitedefaultseppunct}\relax
\EndOfBibitem
\bibitem[Jones(2015)]{Jones2018}
Jones,~R.~O. Density functional theory: Its origins, rise to prominence, and
  future. \emph{Rev. Mod. Phys.} \textbf{2015}, \emph{87}, 897--923\relax
\mciteBstWouldAddEndPuncttrue
\mciteSetBstMidEndSepPunct{\mcitedefaultmidpunct}
{\mcitedefaultendpunct}{\mcitedefaultseppunct}\relax
\EndOfBibitem
\bibitem[Berger \latin{et~al.}(2005)Berger, de~Boeij, and van
  Leeuwen]{JPCC2005}
Berger,~J.~A.; de~Boeij,~P.~L.; van Leeuwen,~R. A physical model for the
  longitudinal polarizabilities of polymer chains. \emph{The Journal of
  Chemical Physics} \textbf{2005}, \emph{123}, 174910\relax
\mciteBstWouldAddEndPuncttrue
\mciteSetBstMidEndSepPunct{\mcitedefaultmidpunct}
{\mcitedefaultendpunct}{\mcitedefaultseppunct}\relax
\EndOfBibitem
\bibitem[Dai(2002)]{Dai2002}
Dai,~H. Carbon nanotubes: opportunities and challenges. \emph{Surface Science}
  \textbf{2002}, \emph{500}, 218 -- 241\relax
\mciteBstWouldAddEndPuncttrue
\mciteSetBstMidEndSepPunct{\mcitedefaultmidpunct}
{\mcitedefaultendpunct}{\mcitedefaultseppunct}\relax
\EndOfBibitem
\bibitem[de~Assis and Dall'Agnol(2019)de~Assis, and Dall'Agnol]{JVSTB2019}
de~Assis,~T.~A.; Dall'Agnol,~F.~F. Minimal domain size necessary to simulate
  the field enhancement factor numerically with specified precision.
  \emph{Journal of Vacuum Science \& Technology B} \textbf{2019}, \emph{37},
  022902\relax
\mciteBstWouldAddEndPuncttrue
\mciteSetBstMidEndSepPunct{\mcitedefaultmidpunct}
{\mcitedefaultendpunct}{\mcitedefaultseppunct}\relax
\EndOfBibitem
\bibitem[Soler \latin{et~al.}(2002)Soler, Artacho, Gale, Garc{\'{\i}}a,
  Junquera, Ordej{\'{o}}n, and S{\'{a}}nchez-Portal]{Soler2002}
Soler,~J.~M.; Artacho,~E.; Gale,~J.~D.; Garc{\'{\i}}a,~A.; Junquera,~J.;
  Ordej{\'{o}}n,~P.; S{\'{a}}nchez-Portal,~D. The {SIESTA} method forab
  initioorder-Nmaterials simulation. \emph{Journal of Physics: Condensed
  Matter} \textbf{2002}, \emph{14}, 2745\relax
\mciteBstWouldAddEndPuncttrue
\mciteSetBstMidEndSepPunct{\mcitedefaultmidpunct}
{\mcitedefaultendpunct}{\mcitedefaultseppunct}\relax
\EndOfBibitem
\bibitem[Perdew \latin{et~al.}(1996)Perdew, Burke, and Ernzerhof]{PBE}
Perdew,~J.~P.; Burke,~K.; Ernzerhof,~M. Generalized Gradient Approximation Made
  Simple. \emph{Phys. Rev. Lett.} \textbf{1996}, \emph{77}, 3865--3868\relax
\mciteBstWouldAddEndPuncttrue
\mciteSetBstMidEndSepPunct{\mcitedefaultmidpunct}
{\mcitedefaultendpunct}{\mcitedefaultseppunct}\relax
\EndOfBibitem
\bibitem[dos Santos \latin{et~al.}(2017)dos Santos, Mota, Rivelino, and
  Gueorguiev]{RB}
dos Santos,~R.~B.; Mota,~F. d.~B.; Rivelino,~R.; Gueorguiev,~G.~K.
  Electric-Field Control of Spin-Polarization and Semiconductor-to-Metal
  Transition in Carbon-Atom-Chain Devices. \emph{The Journal of Physical
  Chemistry C} \textbf{2017}, \emph{121}, 26125--26132\relax
\mciteBstWouldAddEndPuncttrue
\mciteSetBstMidEndSepPunct{\mcitedefaultmidpunct}
{\mcitedefaultendpunct}{\mcitedefaultseppunct}\relax
\EndOfBibitem
\bibitem[Troullier and Martins(1991)Troullier, and Martins]{TM1991}
Troullier,~N.; Martins,~J.~L. Efficient pseudopotentials for plane-wave
  calculations. \emph{Phys. Rev. B} \textbf{1991}, \emph{43}, 1993--2006\relax
\mciteBstWouldAddEndPuncttrue
\mciteSetBstMidEndSepPunct{\mcitedefaultmidpunct}
{\mcitedefaultendpunct}{\mcitedefaultseppunct}\relax
\EndOfBibitem
\bibitem[Dresselhaus and Avouris(2001)Dresselhaus, and Avouris]{Dresselhaus}
Dresselhaus,~D.~G.,~M.~S.; Avouris,~P. \emph{Carbon Nanotubes Synthesis,
  Structure, Properties, and Applications}; Topics in Applied Physics;
  Springer, 2001\relax
\mciteBstWouldAddEndPuncttrue
\mciteSetBstMidEndSepPunct{\mcitedefaultmidpunct}
{\mcitedefaultendpunct}{\mcitedefaultseppunct}\relax
\EndOfBibitem
\bibitem[Kokkorakis \latin{et~al.}(2002)Kokkorakis, Modinos, and
  Xanthakis]{Xanthakis}
Kokkorakis,~G.~C.; Modinos,~A.; Xanthakis,~J.~P. Local electric field at the
  emitting surface of a carbon nanotube. \emph{Journal of Applied Physics}
  \textbf{2002}, \emph{91}, 4580--4584\relax
\mciteBstWouldAddEndPuncttrue
\mciteSetBstMidEndSepPunct{\mcitedefaultmidpunct}
{\mcitedefaultendpunct}{\mcitedefaultseppunct}\relax
\EndOfBibitem
\bibitem[Read and Bowring(2004)Read, and Bowring]{RBowring}
Read,~F.; Bowring,~N. Field enhancement factors of random arrays of carbon
  nanotubes. \emph{Nuclear Instruments and Methods in Physics Research Section
  A: Accelerators, Spectrometers, Detectors and Associated Equipment}
  \textbf{2004}, \emph{519}, 305 -- 314, Proceedings of the Sixth International
  Conference on Charged Particle Optics\relax
\mciteBstWouldAddEndPuncttrue
\mciteSetBstMidEndSepPunct{\mcitedefaultmidpunct}
{\mcitedefaultendpunct}{\mcitedefaultseppunct}\relax
\EndOfBibitem
\bibitem[Zeng \latin{et~al.}(2009)Zeng, Fang, Liu, Yuan, Yang, Guo, Wang, Liu,
  and Zhao]{ZENG2009}
Zeng,~W.; Fang,~G.; Liu,~N.; Yuan,~L.; Yang,~X.; Guo,~S.; Wang,~D.; Liu,~Z.;
  Zhao,~X. Numerical calculations of field enhancement and field amplification
  factors for a vertical carbon nanotube in parallel-plate geometry.
  \emph{Diamond and Related Materials} \textbf{2009}, \emph{18}, 1381 --
  1386\relax
\mciteBstWouldAddEndPuncttrue
\mciteSetBstMidEndSepPunct{\mcitedefaultmidpunct}
{\mcitedefaultendpunct}{\mcitedefaultseppunct}\relax
\EndOfBibitem
\bibitem[Roveri \latin{et~al.}(2016)Roveri, Sant'Anna, Bertan, Mologni, Alves,
  and Braga]{Unicamp2016}
Roveri,~D.; Sant'Anna,~G.; Bertan,~H.; Mologni,~J.; Alves,~M.; Braga,~E.
  Simulation of the enhancement factor from an individual 3D hemisphere-on-post
  field emitter by using finite elements method. \emph{Ultramicroscopy}
  \textbf{2016}, \emph{160}, 247 -- 251\relax
\mciteBstWouldAddEndPuncttrue
\mciteSetBstMidEndSepPunct{\mcitedefaultmidpunct}
{\mcitedefaultendpunct}{\mcitedefaultseppunct}\relax
\EndOfBibitem
\bibitem[Edgcombe \latin{et~al.}(2019)Edgcombe, Masur, Linscott,
  Whaley-Baldwin, and Barnes]{Edgcombe2019}
Edgcombe,~C.; Masur,~S.; Linscott,~E.; Whaley-Baldwin,~J.; Barnes,~C. Analysis
  of a capped carbon nanotube by linear-scaling density-functional theory.
  \emph{Ultramicroscopy} \textbf{2019}, \emph{198}, 26 -- 32\relax
\mciteBstWouldAddEndPuncttrue
\mciteSetBstMidEndSepPunct{\mcitedefaultmidpunct}
{\mcitedefaultendpunct}{\mcitedefaultseppunct}\relax
\EndOfBibitem
\bibitem[Zheng \latin{et~al.}(2004)Zheng, Chen, Li, Deng, and Xu]{LiPRL}
Zheng,~X.; Chen,~G.; Li,~Z.; Deng,~S.; Xu,~N. Quantum-Mechanical Investigation
  of Field-Emission Mechanism of a Micrometer-Long Single-Walled Carbon
  Nanotube. \emph{Phys. Rev. Lett.} \textbf{2004}, \emph{92}, 106803\relax
\mciteBstWouldAddEndPuncttrue
\mciteSetBstMidEndSepPunct{\mcitedefaultmidpunct}
{\mcitedefaultendpunct}{\mcitedefaultseppunct}\relax
\EndOfBibitem
\bibitem[Peng \latin{et~al.}(2008)Peng, Li, He, Chen, Wang, Deng, Xu, Zheng,
  Chen, Edgcombe, and Forbes]{JAPForbes}
Peng,~J.; Li,~Z.; He,~C.; Chen,~G.; Wang,~W.; Deng,~S.; Xu,~N.; Zheng,~X.;
  Chen,~G.; Edgcombe,~C.~J.; Forbes,~R.~G. The roles of apex dipoles and field
  penetration in the physics of charged, field emitting, single-walled carbon
  nanotubes. \emph{Journal of Applied Physics} \textbf{2008}, \emph{104},
  014310\relax
\mciteBstWouldAddEndPuncttrue
\mciteSetBstMidEndSepPunct{\mcitedefaultmidpunct}
{\mcitedefaultendpunct}{\mcitedefaultseppunct}\relax
\EndOfBibitem
\bibitem[Onida \latin{et~al.}(2002)Onida, Reining, and Rubio]{Onida}
Onida,~G.; Reining,~L.; Rubio,~A. Electronic excitations: density-functional
  versus many-body Green's-function approaches. \emph{Rev. Mod. Phys.}
  \textbf{2002}, \emph{74}, 601--659\relax
\mciteBstWouldAddEndPuncttrue
\mciteSetBstMidEndSepPunct{\mcitedefaultmidpunct}
{\mcitedefaultendpunct}{\mcitedefaultseppunct}\relax
\EndOfBibitem
\end{mcitethebibliography}
\end{document}